\documentstyle[11pt,newpasp,twoside,epsf]{article}
\def\edcomment#1{\iffalse\marginpar{\raggedright\sl#1\/}\else\relax\fi}
\reversemarginpar

\begin{document}
\title{Rotation, Statistical Dynamics and Kinematics of Globular
 Clusters} 
\author{D. Lynden-Bell} 
\affil{The Institute of Astronomy, Madingley Road, Cambridge, CB3 0HA.
 dlb@ast.cam.ac.uk, (written at IUCAA, Pune).}
\begin{abstract}
Evolution with mass segregation and the evolution of the rotation of
cores are both discussed for self-similar core collapse.  Evolution
with $\Omega_0\propto \rho_0^{1/2}$ is predicted.  On the
Dynamical Main Sequence of globular clusters the energy emission
from binaries balances the energy expended in expanding the halo.

Newton's exactly solved N-body problem is then given, along with
recent generalisations, all of which have no violent relaxation, but a
new type of statistical equilibrium is discussed.  

Finally, we set the creation of streams in the Galaxy's halo in the
historical context of their discovery.
\end{abstract}
\section{Introduction}
When, over ten years ago, the main focus of my work moved away from
stellar dynamics, three outstanding problems were left unsolved and my
efforts to interest students in these have not borne fruit.  In the
hope that others may be stimulated these questions are posed here.

Q1.  Is it possible to have self-similar core collapse when there is a
continuous distribution of stellar masses?

Q2.  For a weakly rotating cluster is there a self-similar collapse of
the core and how does its rotation evolve as the core radius,
$r_c$, decreases?

Of course question 2 can be tackled for equal mass stars and then one
can make it more realistic by combining both questions together in
Q1 + Q2.

Q3.  It is now widely accepted that just as in stars the energy
generated by nuclear reactions delays thermal evolution, so in star
clusters the energy generated by binary stars delays core-collapse.
Eddington's theory of the Main Sequence led to many observational
consequences.  Can we make a theory of the Dynamical Main Sequence of
globular clusters with real predictive power as to the form these
clusters should take at equilibrium?

Section 2 outlines possible solutions but full solutions involve
multidimensional problems in four or five dimensions (e.g., Energy,
Angular Momentum, Time and Mass).  

Some years ago Inagaki and I (1990) showed that such problems could be
tackled analytically using trial functions and a Local Variational
Principle.  I still think that method should have a great future in
these problems but those wishing to use it should contact Takahashi
(1992,3,5,6) who developed this approach numerically for 3 dimensional
problems, but has since turned to another method.

Since few workers in N-body dynamics have even heard of Newton's exact
solution of an N-body problem for all N and all initial conditions, it
is given in section 3 along with recent developments of this
idea. These problems lead to some pretty dynamics which illustrate
that there {\bf are} exceptions with no violent relaxation and that
thermodynamic equilibria can exist in a system undergoing a macroscopic
radial pulsation or expansion.

Finally in section 4 we discuss the history of the discovery of tidal
streams and give the methods recently invented for finding new ones.
Once some globular cluster extensions or proper motions are reliably
determined these methods will become far more powerful.
\section{Core Collapse, Mass-Segregation, Rotation and the Dynamical
Main Sequence}
Much of globular cluster evolution follows from careful dimensional
analysis, see Inagaki \& Lynden-Bell (1983).  Returning to our
Question 1 we look for a self-similar solution with the density in the
core and inner halo of the form
\begin{displaymath}\rho(r,t)=\rho_0(t)\rho_\ast\left (r_\ast\right )\ \ \ {\rm where}\ \ \ r_\ast = r/r_c(t)\end{displaymath} Here $r_c(t)$ is
the core's radius.

Since the relaxation is far more rapid in the core than in the outer
parts we may set $\partial /\partial t \approx 0$ for large $r$.
\begin{displaymath}0=\left (\partial\ell n\rho /\partial t \right
)_r = d\ell n\rho_0/dt - \left (d\ell n\rho_\ast /dr_\ast
\right ) d\ell nr_c/dt\end{displaymath}
Therefore for large $r$
\begin{displaymath}d\ell n\rho_0 / d\ell nr_c = d\ell n
\rho_\ast / d\ell n r_\ast = - \alpha\end{displaymath} 

Since the expression on the left is a function of $t$ alone, while
that on the right is a function of $r_\ast$ alone, it follows that
both are constant (at a value we have called $-\alpha$).  Thus for all
time \begin{equation}\rho_0 = A r_c^{-\alpha}\end{equation} and
at large $r_\ast$, $\rho_\ast = r_\ast^{-\alpha}$.

The evolution of the core is due to the relaxation so

\begin{equation}d\ell n \rho_0 /dt = K_1/T_c \end{equation}
where $K_1$ is a dimensionless constant and $T_c$ is the relaxation
time in the core.  A standard evaluation of $T_c$ gives for a
velocity dispersion $\sigma$, Binney \& Tremaine (1987)

\begin{equation}T_c^{-1} = 3G^2m\rho \ell n \wedge /\sigma^{3} \end{equation}
and we would like to evaluate this for the core when there is a
distribution of stellar masses.  It would seem natural to take the
distribution function to be an equilibrium Maxwell-Boltzmann there
proportional to $\exp \left [-\beta m \left ({1\over 2} v^2 - \psi_0\right
)\right ]$, but this cannot be true at energies close to the escape
energy.  The closeness of the escape energy is well emphasised by the
following argument from the Virial Theorem.  
\begin{equation}2T=\sum_I m_Iv_I^2 = - V = {1 \over 2} \sum_I
m_I\psi_I \end{equation}  
Now $\psi_I = {1\over 2} v^2_e$ where $v_e$ is the velocity of escape to
infinity, hence for any cluster obeying the virial theorem
\begin{equation}\langle mv^2 \rangle = {1 \over 4} \langle mv^2_e
\rangle \ \ . \end{equation}
So that the mass weighted $r ms$ velocity is HALF the mass weighted
$r ms$ velocity of escape.  At a thermodynamic equilibrium we
have equipartition of ${1\over 2} mv^2$ so the above expression implies that
stars with less than a quarter of the mean mass have an rms velocity
equal to the escape velocity.  This cannot occur in practice.  Thus
the equipartition of the Maxwell-Boltzmann distribution must be
modified and as Michie (1963) first showed the lowered Maxwellian of
the Michie-King models [evaluated in full by King (1966)] does this
approximately viz
\begin{equation}f=B(m)\left \{\exp \left [-\beta m \left ({1 \over
2}v^2-\psi \right )\right ] -\exp \left (\beta m \psi_e \right )\right
\}\end{equation} 
Although in well developed core collapse such models have insufficient
temperature contrast between the centre and the halo, Lynden-Bell \&
Wood (1968), Lynden-Bell \& Eggleton (1980), nevertheless we shall
adopt such a form for the low energy stars in the core (for which
relaxation is the most rapid).  If $F(m)dm$ gives the number density
of stars at mass $m$ in the core, so that
\begin{equation}\rho_0 = \int m F(m)dm\end{equation}
then we may re-express the relaxation rate in terms of the central
`temperature' $\beta^{-1}=\sigma^2/m$ and obtain from (3)
\begin{equation}T_c^{-1}=3G^2\beta^{3/2}\rho_0 \ell n \wedge \langle
m^{7/2} \rangle / \langle m \rangle \end{equation}
where
\begin{equation}\langle m^{7/2} \rangle = \int m^{7/2} F(m)dm/\int
F(m)dm \end{equation}

Evidently the relaxation rate, $T^{-1}_c$, which determines the
overall evolution of the cluster, depends on the mean seven halves
power of the mass, evaluated over the core of the cluster.  As
evolution proceeds, the lighter stars in the core are preferentially
expelled from it but simultaneously the heavy stars are gradually
eliminated via stellar evolution.  Before using (8) in (2) we need to
re-express $\beta$ in terms of $\rho_0,r_c$ and $\langle m
\rangle$.  If $M_c={4 \over 3}\pi \rho_0r^3_c$ is the core mass
then  
\begin{equation}3 \sigma^2=3(\beta \langle m \rangle
)^{-1}=GM_c/r_c={4 \pi \over 3} G \rho_0r^2_c\end{equation}
where we have chosen to define $r_c$ so that the constant of
proportion is one.  
\begin{equation}T^{-1}_c=K_2G^{1/2}\rho_0^{-1/2}r_c^{-3}m_c=K_2G^{1/2}
A^{-1/2}m_c (\rho_0/A)^{(6-\alpha)/(2\alpha)}\end{equation} 
where we have used (1) to express $r_c$ in terms of $\rho_0$, $K_2$ is
the dimensionless constant $\left ({81  \over 8 \pi ^{3/2}} \ell n
\wedge \right )$ and $m_c(t)=\langle m^{7/2}\rangle /\langle m \rangle
^{5/2}$.  To use expression (11) to solve (2) we must first evaluate
the time dependence of $m_c$.  This is caused by two separate effects,
the progressive expulsion of lower mass stars from the core increases
$m_c$ but it is mainly dependent on the high masses and they diminish
steadily as stellar evolution and stellar death take their toll.  With
such a high power of the mass involved it is likely that stellar
evolution plays the dominant role with $m_c$ behaving similarly to the
age cut off.  This suggests the approximate form 
\begin{displaymath}m_c=m_\odot \left (t/t_\odot \right )^{\delta
-1}\ {\rm where}\ \delta \simeq 2/3,\ t_\odot \simeq 10^{10} {\rm yr},\
 t>10^6 {\rm yr} \end{displaymath} 
and $t$ is time measured from the time the cluster was created.
 Inserting this into (11) and (11) into (2) the resulting equation for
 $\rho_0$ is readily solved to give 
\begin{equation} \rho_0 = C \vert t^\delta_0-t^\delta\vert^{-2\alpha
 /(6-\alpha)}\end{equation}
and therefore
\begin{equation}r_c=(A/C)^{1/\alpha}\vert t^\delta_c-t^\delta
\vert^{2/(6-\alpha )}\end{equation}
where $t_c$ is a constant of integration which gives the time of core
collapse and the constant $C$ is
\begin{displaymath}C=A \left [{6-\alpha \over
2\alpha\delta}K_2G^{1/2}A^{-1/2}m_\odot t_\odot^{1-\delta} \right
]^{-2\alpha / (6-\alpha)}\ . \end{displaymath}
As in Inagaki \& Lynden-Bell (1983), we expect (12) and (13) to hold
also after core collapse but then $\rho_0$ becomes a
characteristic density rather than the central one which remains
infinite if we ignore binaries and the giant gravothermal
oscillations.  

Of course in the absence of stellar evolution $m_c$ will increase and
the \linebreak parameterisation 
\begin{displaymath}m_c=m_\odot(t/T_c)^\delta\end{displaymath}
might then be more appropriate but though such a model is soluble I
shall not detail it here.

Expressions (12) and (13) can only be considered the solution to our
problem once $\alpha $ is known.  As Lynden-Bell and Eggleton showed
$\alpha $ emerges as an eigen value in the full theory and for the
equal mass case $\alpha = 2.22$.  It is readily seen that $\alpha$
must lie between the equal mass isothermal sphere with $\alpha = 2$
and the limiting case with infinite core binding energy $\alpha =
5/2$.  With the heavy masses more concentrated to the core one may
expect $2.22 <\alpha <2.5$ and perhaps toward the upper end of that
range, on the other hand once most of the lighter stars have been
ejected from the core one expects the evolution to revert toward the
single mass case.  Our best {\bf guess} is therefore 
\begin{equation}\alpha = 2.3 \pm 0.08 \ . \end{equation}

In discussing Question 2 we note that the flattening of a cluster is
of second order in $\Omega$ the angular velocity and may be neglected
to first order.  In any potential of the form
\begin{equation}\psi =\psi_0(r)+r^{-2}\psi_2(\theta)\end{equation}
the expression $I={1\over 2}J^2-\psi_2$ is an exact integral of the
motion where ${\bf J} = ({\bf r} \times {\bf v})$.  In practice any
non-spherical part of $\psi$ is second order in $\Omega$ and the
motion of a star is well approximated as lying in a precessing plane.
The rate of precession can be worked out from the couple that the
non-spherical potential exerts on the unperturbed rosette orbit that
the star would have in the absence of asphericity.  Thus in practice
$\vert {\bf J}\vert $ is almost an integral of the motion while $J_z$
and the energy, $\epsilon$, are exact ones in the absence of
evolution.  In practice it is convenient to use the radial action
$J_r={1 \over 2\pi }\oint {{\sqrt{2[\epsilon +
\psi_0(r)]-J^2/r^2}}}dr$ in place of $\epsilon$.  In potentials of the
form (15) it is an exact integral but for more general forms of $\psi$
it is only approximate.  A particular advantage of using $\vert {\bf
J}\vert ,\ J_r, \ J_z$ is that they are adiabatically invariant for
slow changes in the potential $\psi$.  Thus if we express the
distribution function of a rotating cluster in the form $f=f(\vert
{\bf J}\vert,\ J_r,\ J_z,\ t)$ where $f$ evolves little in one orbital
time, then ${\bf \partial} f/\partial t$ is $(\partial f /\partial t)\
_{\rm{encounters}}$ and there is no supplementary term due to the
change of global potential induced by these encounters since for such
changes the $J$ (although not $\epsilon$) are adiabatically invariant.
The $J$ have dimensions of $\sigma r_c$ and $f$ has the dimensions of
$\rho_0\sigma^{-3}$, so in self-similar evolution the evolving $f$
will have the form
\begin{equation}f=\rho_0\sigma^{-3}F_\ast\left (\vert {\bf J}^\ast\vert ,\
J^\ast_r,\ J^\ast_z \right )\end{equation}
where $F_\ast $ is a dimensionless function of its dimensionless
arguments \linebreak $\vert {\bf J}^\ast \vert = \vert {\bf J}\vert / (\sigma
r_c),\ J^\ast_r = J_r/(\sigma r_c),\ J^\ast_z=J_z /(\sigma r_c). $  As
before $\rho_0 = Ar_c^{-\alpha}$ so $\sigma r_c\propto
r^{2-\alpha/2}_c$.  These same principles may be applied to discuss
the way cluster core rotation evolves as the core contracts.  
\vspace{-1pt}
Near the centre, relaxation is quite fast so we expect a rotating
Maxwellian at energies little affected by escape 
\begin{equation}f = B\exp - \sigma^{-2}(\epsilon - \Omega_0J_z)\ \ . \end{equation}
Our question is how does $\Omega_0$ evolve during core collapse.  In
self-similar evolution the quantities that are dimensionless do not
evolve, so $\Omega_0 r_c/\sigma = {\rm const} $.  Hence as the core
shrinks we have our prime result
\begin{equation}\Omega_0\propto r^{-\alpha /2}_c \ \ \ \ \ \ \ \
\alpha \simeq 2.22 \ . \end{equation} So as $r_c$ shrinks, the core
should rotate faster and faster.  This result is very sensibly between
the $\Omega_0r^2_c = {\rm const}$ that would follow if the heat flowed
out of the core much more readily than angular momentum, and the
$\Omega = {\rm const}$ that would follow if the opposite were true.
Away from the centre it is plausible that the system remembers the
rotation it had when that part of the inner halo was part of the core;
in reality it gains a little more angular momentum as it leaves the
remaining core, because the remaining core loses it.  This suggests
the behaviour $\Omega \sim r^{-\alpha/2}$ but since $\Omega^2$ and
$G\rho$ have the same dimensions a better prediction is that $\Omega$
behaves approximately as $\rho^{1/2}$; hence we conclude that cluster
cores and inner haloes rotate as
\begin{equation}\Omega =
\Omega_0(\rho/\rho_0)^{1/2}=\Omega_0\rho_\ast^{1/2}\end{equation} with
$\Omega_0$ evolving according to (18).  For $Q_1 + Q_2$ the same
results will hold with $\alpha$ given by (14).  Such predictions
should be compared with the numerical work of Spurzem (2000) and the
rapidly growing data from observations.  A further consequence of (19)
is that $\Omega^2/\pi G\rho$ which determines the flattening will be
constant in the core and inner halo thus the isophotes there should
have $b/a \ {\rm constant }$ whereas if the core evolved with
$\Omega_0 \propto r^{-2}_c$ the central isophotes would be more
flattened.  By contrast a uniformly rotating cluster has the greatest
flattening in the outer isophotes.  Turning now to Question 3 the
theory of the Dynamical Main Sequence of Globular Clusters as yet
awaits someone brave enough to make strong hypotheses such as `The
only binaries that matter are in the core and the energy flux through
the inner halo is constant and drives either the escape from the
cluster or the expansion of its halo.'
\vspace{-1pt}
Of course if core collapse ceases due to binaries then the steady core
will relax to uniform rotation which will gradually extend from the
core outwards.
\section{Newton's N-body Problem and its Generalisations}
Let the force on body $I$ due to body $J$ be of the form 
\begin{equation}{\bf F}_{IJ} = k m_I m_J ({\bf r}_J - {\bf
r}_I)\end{equation} 
We sum over {\bf all} $J$ to get the force on $I$ since the $I=J$ term
is zero.
\begin{displaymath}{\bf F}_I = km_I(\sum m_Jr_J-Mr_I)=km_IM({\bf
\bar{r}}-{\bf r}_I)\end{displaymath} 
where $M$ is the total mass.
Thus the total force is directed to the centre of mass ${\bf {\bar
r}}$ and is proportional to the distance from it.
\vspace{-1pt}
Newton's (1687) general solution is that each particle moves in a
central ellipse about the centre of mass which itself moves uniformly
in a straight line.  Newton's system has a total potential energy
\begin{equation}V+{1\over 2}KM^2r^2\end{equation}
where
\begin{equation}r^2=M^{-1}\sum m_I({\bf r}_J-{\bf{\bar
r}})^2\end{equation}
The Virial theorem reads 
\begin{equation}{1 \over 2}\ddot{I}=2T+nV\end{equation}
where $V \propto r^{-n}$ so $n=-2$ for the above system.
\vspace{-1pt}
To generalise.  Newton's result consider systems with total potential
energies of the form
\begin{displaymath}V=V(r)\end{displaymath}
with $r$ given by (22).  This problem has a singular beauty if we
consider the 3N dimensional space in which the ${\bf X}_I = {\sqrt
{m_I \over M}}({\bf r}_I-{\bf{\bar r}})$ are the coordinates.
\vspace{-1pt}
Letting the 3N vector ${\bf r} = ({\bf x}_1, {\bf x}_2 \cdots , {\bf
x}_N)$ we note that $r^2$ is given by (22).  The initial conditions in
this space are the initial values of ${\bf r}$ and ${\bf{\dot r}}$.
The accelerations in this space are central since $V=V(r)$ so the
acceleration also lies in the plane defined by the initial ${\bf r}$
and ${\bf{\dot r}}$.  Thus the motion continues in that plane.  In
fact the whole problem now reduces to the planar orbit problem under
the action of the central potential $V(r)$.  Back in 3 space each
particle feels the central force $V^\prime (r)m_IM^{-1}({\bf r}_I-{\bf
\bar r})/r$ so each particle orbits in a plane.  Its motion may be
obtained by projecting the motion of the representative point in 3N
down into the plane of the motion of particle I.
\vspace{-1pt}
As $r$ changes periodically around the planar orbit in 3N space it
follows that $Mr^2$ vibrates forever so there is no violent
relaxation to a Virial equilibrium.  
\vspace{-1pt}
One may generalise this result still further by taking
$V=V_0(r)+r^{-2}V_2({\bf r}/r)$ then the Virial theorem reads
\begin{displaymath}{1 \over 2}
Md^2r^2/dt^2=2T-rV^\prime_0(r)+2V_2=2E-2V_0-rV^\prime_0\end{displaymath} 
since this last expression does not involve $V_2$ it follows that $r$
vibrates in the same way as it did when $V_2 $ was 0.  Hence once
again there is no violent relaxation.  Pretty N-body problems of this
type are given by the forces
\begin{displaymath}F_{IJ}=Gm_Im_J({\bf r}_J-{\bf r}_I)\left [{1 \over
r^3}-{k\over ({\bf r}_J-{\bf r}_I)^4}\right ]\end{displaymath}
where $r$ is given by (22) or alternatively
\begin{displaymath}F_{IJ}=G^\prime m_Im_J({\bf r}_J-{\bf r}_I)\left [1-k({\bf
r}_J-{\bf r}_I)^{-4}\right ]\end{displaymath}
both of these force laws have long range attractions and short range
repulsions so they may produce liquid-like and solid-like phases .  In
each case the vibration in $r$ separates from the other motions so if
lattices are possible, lattices with a breathing pulsation will be
too.  It is possible to do a statistical mechanics of the `angular'
motions only while $r$ remains breathing but for details the reader
should refer to Lynden-Bell \& Lynden-Bell (1999a).  The 1999b paper
solves Schrodinger's equation for both spin 0 Bosons and spin ${1
\over 2}$ Fermions giving the only known non-trivial three dimensional
N-body solutions for interacting Bose or Fermi gases.  The degenerate
Fermi case gives a white-dwarf-like solution.
\vspace{-1pt}
\section{Streams 1950-2000}
Streams about the Galaxy have much in common with the meteor streams
left behind by comets in the solar system which were discovered
earlier.  However the galactic orbits are rosettes rather than
ellipses and their planes precess about the galactic pole due to the
aspherical potential.  Bertil Lindblad (1958)(1961) was one of the
first to consider the theory of streams spreading from a common origin
but already Eggen (1959) (1989) was on the march finding moving groups
in the velocity space among both disc and halo stars.  As data were
refined Eggen changed his mind about the central velocities of some
groups and this led to consequent changes in their membership which
only increased scepticism among his critics.  However Eggen was
convinced he was onto something and claimed that it was much easier
for others to criticise the inclusion of a star when new data showed
its membership to be unlikely, than to discover the group to start
with.

\vspace{-1pt}
The subject of streams really come to life with the discovery of the
Magellanic Stream stretching more than 120$\deg $ around the galaxy.
But this discovery was the culmination of a sequence.  In 1965 Nan
Dieter discovered a high velocity cloud at the South Galactic Pole.
Soon afterward Hulsbosch \& Raimond (1966) published their survey of
such objects and showed that cloud to be strongly elongated.  It was
the technological innovation of Wrixon's low noise diodes that allowed
Wannier \& Wrixon (1972) to see fainter 21 cm emission.  Their great
arc of 21 cm emission through Dieter's cloud showed a systematic
variation of radial velocity by several hundred km/sec across the sky.
At Herstmonceux Kalnajs (1972) excitedly drew my attention to their
paper claiming that it must be of Galactic scale.  A day or two later
he discovered that both the line of the arc and the radial velocity
extrapolated to the Magellanic Clouds.  Mathewson too was very excited
by this discovery and no sooner had he got back to Australia than he,
Cleary \& Murray (1974) started delineating how the stream joined the
Magellanic Clouds.  A few years later during a non-photometric night
at Sutherland I was imagining how the Magellanic Stream lay across the
sky when it struck me that other satellites of the Milky Way might be
members of streams and might likewise have high velocity clouds
associated with them.  I rapidly found that Draco \& Ursa Minor are
almost antipodal to the LMC and the SMC in the galactocentric sky.
This meant that they too might be long lost members of the Magellanic
Stream, Lynden-Bell (1976).  Kunkel \& Demers (1977) noticed the dwarf
spheroidals were distributed non-randomly over the sky but their
compromise plane was some 30$\deg$ from the Magellanic Stream.  In
their attempt to explain the antipody of Draco \& Ursa Minor, Hunter
\& Tremaine (1977) noticed that their elongations lay along the stream
while I searched for further streams using elongations as a
supplementary guide, Lynden-Bell (1982); however it was not until 1995
that this developed into a systematic method.  Since tidal tearing
occurs essentially in the plane of orbit, one needs to find objects
scattered along a great circle in the galactocentric sky.  Such great
circles are hard to see in any of the projections of the sky so we
need a good method of finding them.  An object may be associated with
the set of all great circles that pass through it.  The poles of these
great circles lie on another great circle which we call a polar path.
(The pole of that great circle lies of course at the object we started
with).  The great circle through any {\bf two} objects has its pole at
the intersection of their polar paths.  If 3 objects lie on a great
circle all 3 polar paths will go through the same point.  Thus we may
find objects with common great circles by plotting their polar paths
(in any equal area projection) and looking for multiple intersections.
Figure 1a which plots 45 polar paths for dwarf spheroidals and outer
globular clusters shows a result that more closely resembles a ball of
wool than a useful scientific plot.  More information is needed to get
any definitive groupings.  A known proper motion would reduce the
polar path of an object to a very short segment determined by the
directional error of the proper motion.  If we are prepared to assume
that the elongation of an object is along its orbit, as appears to be
true for Draco \& Ursa Minor and might be true for the outer
extensions recently found around globulars (e.g., Grillmair et. al.,
Meylan this volume) then likewise the pole must lie in only a small
sector of the polar path (see Figure 1b).  But great circles are not
enough.  Radial velocities provide a discriminant.  If all the objects
in a stream follow the orbit of the progenitor then they will share
the same specific energy $E$ and specific angular momentum $h$.

\begin{figure}
\plotone{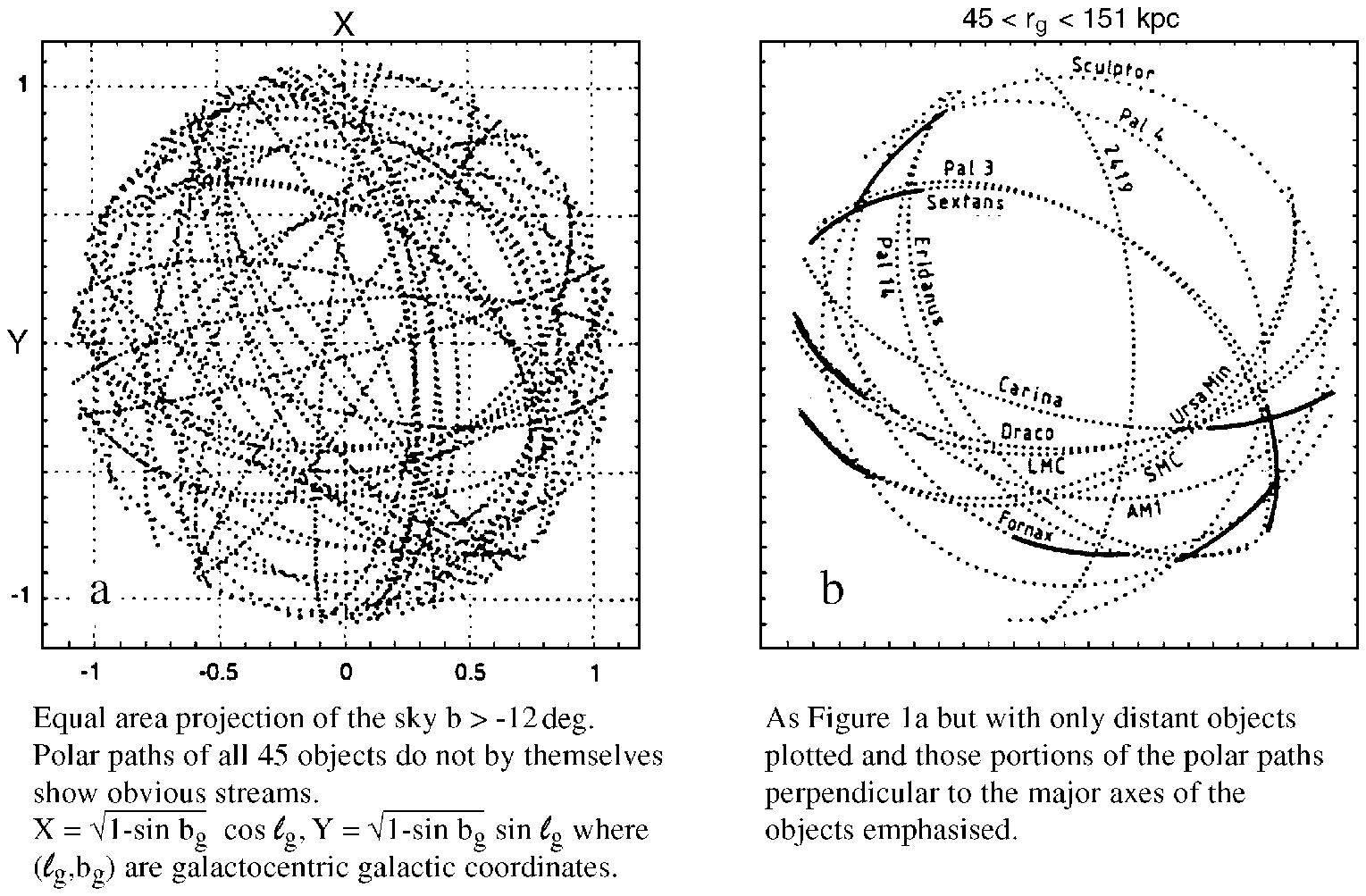}
\caption{a \& b }
\bigskip

\bigskip

\bigskip

\bigskip

\plotone{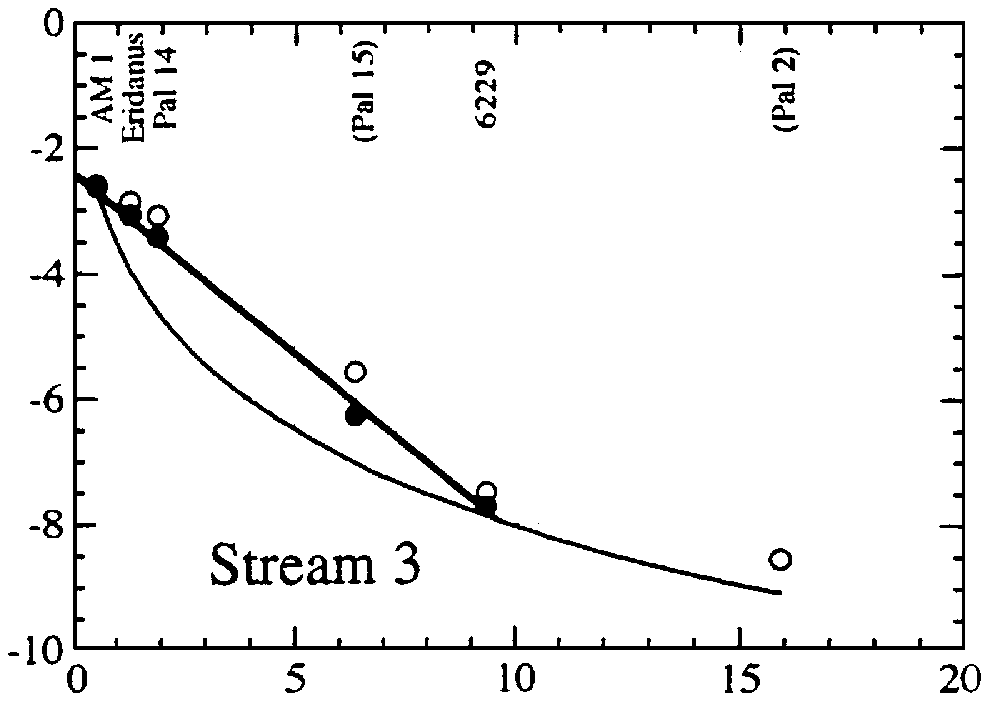}
\caption{The radial energy plotted against $r^{-2}_g$.  The gradient fives
$-h^2/2$ where $h$ is the specific angular momentum of the stream.
From this the predicted proper motions are deduced (see L-B)$^2$ 1995.}
\end{figure}

\newpage
\noindent
Taking sets of objects that share a common
polar intersection and therefore lie on a galactocentric great circle
one may plot $E_r = E - {1 \over 2}h^2r^{-2}={1\over
2}v^2_r-\psi(r)$ against $r^{-2}$ the inverse square of their
galactocentric distances.  This should be  a straight line so
$E$ and $h^2$ can be deduced as the intercept and gradient of
such a line.  In practice the observed radial velocity is not the
galactocentric one so $v_r$ can be deduced by estimating an initial
${\bf h}$ and then iterating to find a consistent gradient of
$-{1\over 2}h^2$.  In practice this method converges rapidly but
needless to say some of the objects do not lie on straight lines and
are interlopers rather than stream members.  Since the method gives
$h^2$ from the observed radial velocities, and since the poles of the
great circle involved are known, we have used these methods not only to
produce possible streams among the dwarf spheroidals and outer
globulars but also to predict their proper motions.  The method works
well for the Magellanic Stream and for the Sagittarius Stream which
was discovered by Ibata, Gilmore \& Irwin (1995).  There is now great
hope that we can identify orbital streams associated with the later
merging events emphasised by Searle and Zinn.  Further work may take
us back to the orbits of objects that took part in the initial
formation itself.  The ELS (1962) picture had the bulk of the metal
poor globulars formed in the fragmentation of the initial collapse,
but following Ambartsumian we know that most of the stars would have
been formed in looser aggregates and stellar associations which
subsequently broke up and spread around the Galaxy.  The idea that ELS
pictured a perfectly smooth collapse is not in their paper but has
been added by others who have not read it.  Eggen (1959) had already
worked on streams in the halo, the fragmentation of collapsing systems
was treated by Hunter (1962), and I was working on the Large Scale
Instabilities of shape that would make the collapse non-axis
symmetric, Lynden-Bell (1964).  The non-linear behaviour of these
instabilities was later treated in the simple non-rotating case by
Lin, Mestel \& Shu (1965).  Those who want to read for themselves what
was thought at that time would do well to read my paper to IAU
Symposium No. 31 on the Formation of the Galaxy, Lynden-Bell (1967).
Streams from the initial collapse are less likely to be preserved than
the younger streams from late additions, but finding correlated
stellar motions from the initial creation of the Galaxy was what ELS
was all about and their method of using abundances and adiabatic
invariants as fossilised evidence on how the Galaxy formed is probably
its most lasting legacy to astronomy.  The merging of galaxies was
suggested as important in the very percipient paper on bridges and
streams by the Toomre's (1972) and White's (1976) demonstration of
fragmentation of a {\bf uniform} overdensity showed that even within
the free fall time of the whole much fragmentation and subsequent merging
of those fragments occurs.

Returning to streams there is too much modern work to detail all of
it, but work by Johnston (1998) holds out the prospect of more
accurate determination of the Galaxy's potential using streams of
stars just escaping from a satellite galaxy.  By contrast, Helmi \&
White (1999) take such stars to be emitted with a not inconsiderable
velocity dispersion so such streams would be harder to detect and
would define the potential less accurately.  It is not yet clear to me
which view is more realistic.  Lastly, Putman et. al. have at last
detected a forward stream from the Magellanic Clouds, albeit somewhat
skew of the main great circle.  Currently the most plausible
explanation is that it was gas around the SMC that was torn but  much
of the forward stream ran into the LMC and the rest was diverted by
it, however other interpretations are possible.

\acknowledgments

The theory of rotation was developed after the conference while
writing up this contribution under the hospitality of IUCAA, Pune,
India.

\end{document}